\documentclass[journal]{IEEEtran}
\usepackage{cite}
\usepackage{amsmath,amssymb,amsfonts}
\usepackage{algorithm}
\usepackage{algorithmic}
\usepackage{graphicx}
\usepackage{textcomp}
\usepackage{xcolor}

\newtheorem{lemma}{Lemma}

\newtheorem{theorem}{Theorem}
\ifCLASSINFOpdf
\else
\fi
\begin{document}
%
\title{FFT-Free PAPR Reduction Methods for OFDM Signals}
%
%
%
\author{Hao Su,
         Jiangtao Wang,~\IEEEmembership{Member,~IEEE}\footnote{Guangzhou Institute of Technology, Xidian University, Guangzhou, 510555, China} and Yongchao Wang,~\IEEEmembership{Senior Member,~IEEE}\\
\thanks{H.~Su and Y.~Wang are with the school of Telecommunication Engineering, Xidian University, Xi'an 710071, China. J.~Wang is with Guangzhou Institute of Technology, Xidian University, Guangzhou, 510555, China.(emai1s: ychwang@mail.xidian.edu.cn, hsu@stu.xidian.edu.cn, jtwang@xidian.edu.cn)}}

\maketitle

\begin{abstract}
In this paper, we propose two low-complexity peak to average power ratio(PAPR) reduction algorithms for orthogonal frequency division multiplexing(OFDM) signals. The main content is as follows: First, a non-convex optimization model is established by minimizing the signal distortion power. Then, a customized alternating direction method of multipliers(ADMM) algorithm is proposed to solve the problem, named T-ADMM along with an improved version called TCU-ADMM. In the algorithms, all subproblems can be solved analytically, and each iteration has linear computational complexity. These algorithms circumvents the challenges posed by repeated fast Fourier transform(FFT) and inverse FFT(IFFT) operations in traditional PAPR reduction algorithms. Additionally, we prove that the T-ADMM algorithm is theoretically guaranteed convergent if proper parameter is chosen. Finally, simulation results demonstrate the effectiveness of the proposed methods.
\end{abstract}

\begin{IEEEkeywords}
Orthogonal frequency division multiplexing(OFDM), peak to average power ratio(PAPR), alternating direction method of multipliers(ADMM).
\end{IEEEkeywords}

%
\IEEEpeerreviewmaketitle

\section{INTRODUCTION}
%
%
%
%
Orthogonal frequency division multiplexing (OFDM) technique offers high spectral efficiency and strong resilience to multipath fading, making it widely used in wireless communication systems, including 4G, 5G and future wireless communication systems\cite{1}\cite{2}. However, the time-domain waveform of OFDM signals exhibits a high peak to average power ratio (PAPR)\cite{3}, which leads to nonlinear distortion when passing through the power amplifier (PA) of a transmitter. Therefore, the PAPR problem should be carefully addressed in OFDM system.

In recent years, numerous approaches have been developed to address the PAPR issue in OFDM signals. These techniques are typically classified into three main categories: coding techniques, multiple signaling and probabilistic techniques, and signal distortion techniques\cite{24}. Coding techniques combine PAPR reduction with some coding schemes such as Gray codes \cite{5}, LDPC codes\cite{6} and Turbo codes\cite{7}. Multiple signaling techniques adopt multiple alternative signal techniques or probability analysis techniques to reduce PAPR. Typical techniques include selective mapping (SLM)\cite{8}, partial transmit sequence (PTS)\cite{9}.

Signal distortion techniques have emerged as a simpler and computationally efficient solution for PAPR problem, offering a good balance between complexity and performance. Common methods include Iterative clipping and filtering (ICF) \cite{12} and companding \cite{13}, etc. Signal distortion methods were also integrated with optimization techniques. The PAPR problem was first formulated as a second order conic programming (SOCP) problem in \cite{15}. Following this, several methods\cite{16}-\cite{18} were proposed to solve PAPR problem. More recently, the PAPR reduction problem combined with alternating direction
method of multipliers (ADMM) was first explored in \cite{21}. Then, an ADMM-based PAPR reduction method with theoretical convergence guarantees was introduced in \cite{22}.

While existing PAPR reduction methods can alleviate the PAPR problem, they still face challenges, especially the computational complexity. Those methods rely on a large number of fast Fourier transform(FFT)/inverse FFT(IFFT) operations, resulting in high computational complexity and posing significant challenges for practical implementation.

In this paper, we propose two FFT-free PAPR reduction methods, significantly reducing computational complexity while maintaining system performance and achieving efficient PAPR reduction. First, We formulate a optimization model based on minimizing signal distortion power. Then, We design two ADMM algorithms, named T-ADMM and TCU-ADMM, to solve the model, where all subproblems can also be solved analytically. Additionally, We prove that the T-ADMM algorithm is theoretically guaranteed convergent if proper parameter is chosen. Finally, simulation results demonstrate the effectiveness of the proposed methods.

\section{PROBLEM FORMULATION}

Consider an OFDM system with $N$ carriers. The frequency-domain OFDM symbol  $\mathbf{s}=[s_1,s_2,\ldots,s_N] \in \mathbb{C}^N$  corresponding to the $\ell$-times oversampled time-domain symbol $\mathbf{x} \in \mathbb{C}^{\ell N}$, can be expressed as

\begin{equation}
  \mathbf{x}=\mathbf{F s},
\end{equation}
 where $\mathbf{F}$ is the first $N$ columns of the $\ell N \times \ell N$ IFFT rotation matrix. Under $\ell$-times oversampling, the PAPR of the OFDM time-domain symbol is defined as

\begin{equation}\label{PAPRdef}
    \operatorname{PAPR}=\frac{\max\limits _{i=1, \ldots, \ell N}\left|x_i\right|^2}{\frac{1}{\ell N} \sum\limits_{i=1}^{\ell N}\left|x_i\right|^2}=\frac{\|\mathbf{x}\|_{\infty}^2}{\frac{1}{\ell N}\|\mathbf{x}\|_2^2}.
\end{equation}

From \eqref{PAPRdef}, it is clear that the time-domain symbols of OFDM suffer from the high PAPR problem. Common PAPR reduction methods require multiple FFT/IFFT operations, leading to high computational complexity. To address this issue, we propose a new method which only involve time-domain signals. Based on a strategy that minimizes signal distortion, we establish the following optimization model

\begin{subequations}\label{PAPRmodel}
  \begin{align}
\underset{\mathbf{u} \in \mathbb{C}^{\ell N}, \mathbf{x} \in \mathbb{C}^{\ell N}}{\operatorname{minimize}} & \frac{1}{2}\|\mathbf{u}\|_2^2,\label{PAPRobj} \\
\text { subject to }  & \frac{\|\mathbf{x}\|_{\infty}^2}{\frac{1}{\ell N}\|\mathbf{x}\|_2^2}=\alpha, \label{PAPR} \\
& \mathbf{x}=\mathbf{x}_o+\mathbf{u},
\end{align}
\end{subequations}
where $\mathbf{u}$ represents the distortion introduced to reduce the PAPR, and $\mathbf{x}_o$ is the original time-domain signal. The PAPR constraint \eqref{PAPR} is non-convex, which makes the problem more challenging to solve. Therefore, we relax the PAPR constraint and reformulate the model as follows

\begin{subequations}\label{model}
\begin{align}
\underset{\mathbf{u} \in \mathbb{C}^{\ell N}, x \in \mathbb{C}^{\ell N}}{\operatorname{minimize}} & \frac{1}{2}\|\mathbf{u}\|_2^2, \label{obj}\\
\text { subject to } & \|\mathbf{x}\|_{\infty} \leq \beta,\label{inf} \\
& \mathbf{x}=\mathbf{x}_o+\mathbf{u},
\end{align}
\end{subequations}
where $\beta=\alpha\sqrt{\frac{1}{\ell N}}\|\mathbf{x}_o\|_2$.

In model \eqref{model}, the proposed model only processes the time-domain signal without involving the frequency-domain signal. This means that the FFT/IFFT operations are not required during the solution process. Since the objective function \eqref{obj} minimizes the energy of signal distortion, $\mathbf{x}$ is optimized to keep as close as possible to the original signal. As a result, the distortion energy mainly concentrates within the same bandwidth as the original signal $\mathbf{x}_o$, which, in a sense, brings some effect to limit out-of-band emissions(OOBE).

\section{PROPOSED ALGORITHM:T-ADMM AND TCU-ADMM}

The alternating direction method of multipliers (ADMM) has emerged as a robust and versatile optimization framework, particularly suited for distributed and large-scale problems \cite{20}. In this section, We design two low-complexity ADMM algorithms, T-ADMM alone with an improved version called TCU-ADMM, to solve the model. In both algorithms, all subproblems have a computational complexity of $\mathcal{O}\left(\ell N\right)$. Furthermore, We prove that the T-ADMM algorithm is theoretically guaranteed convergent if proper parameter is chosen.

\subsection{T-ADMM Algorithm Framework}

 The augmented Lagrangian function for model \eqref{model} can be written as

\begin{equation}
\begin{aligned}
L_\rho\left(\mathbf{u}, \mathbf{x}, \mathbf{y}\right)=&\frac{1}{2}\|\mathbf{u}\|_2^2+\operatorname{Re}\left(\mathbf{y}^H\left(\mathbf{x}-\mathbf{x}_o-\mathbf{u}\right)\right)\\
&+\frac{\rho}{2}\left\|\mathbf{x}-\mathbf{x}_o-\mathbf{u}\right\|_2^2,
\end{aligned}
\end{equation}
where $\mathbf{y} \in \mathbb{C}^{\ell N}$ is the Lagrange multiplier and $\rho>0$ is the penalty parameter. The proposed T-ADMM algorithm is shown as follows

\begin{subequations}
\begin{align}
& \mathbf{u}^{k+1}=\arg \min L_\rho\left(\mathbf{u}, \mathbf{x}^k, \mathbf{y}^k\right), \label{sub1}\\
& \mathbf{x}^{k+1}=\underset{\mathbf{x} \in \mathcal{X}}{\arg \min } L_\rho\left(\mathbf{u}^{k+1}, \mathbf{x}, \mathbf{y}^k\right), \label{sub2}\\
& \mathbf{y}^{k+1}=\mathbf{y}^k+\rho\left(\mathbf{x}^{k+1}-\mathbf{x}_o-\mathbf{u}^{k+1}\right),\label{lag}
\end{align}
\end{subequations}
where $\mathbf{x} \in \mathcal{X}$ denotes constraint \eqref{inf}. The key to solving the ADMM algorithm lies in efficiently solving subproblems \eqref{sub1} and \eqref{sub2}. Based on the augmented Lagrangian function, the first subproblem \eqref{sub1} can be equivalently written as

\begin{equation}\label{sub1model}
\underset{\mathbf{u}\in \mathbb{C}^{\ell N}}{\operatorname{minimize}} \frac{1}{2}\|\mathbf{u}\|_2^2-\operatorname{Re}\left(\mathbf{y}^H \mathbf{u}\right)+\frac{\rho}{2}\left\|\mathbf{x}^k-\mathbf{x}_o-\mathbf{u}\right\|_2^2.
\end{equation}

Since subproblem \eqref{sub1model} is an unconstrained second-order optimization problem with variable $\mathbf{u}$, its optimal solution satisfies $\nabla_{\mathbf{u}} L_\rho\left(\mathbf{u}, \mathbf{x}^k, \mathbf{y}^k\right)=0$, i.e.,

\begin{equation}
\nabla_{\mathbf{u}} L_\rho\left(\mathbf{u}, \mathbf{x}^k, \mathbf{y}^k\right)=\mathbf{u}-\mathbf{y}^k-\rho\left(\mathbf{x}^k-\mathbf{x}_o-\mathbf{u}\right)=0.
\end{equation}

Then, we can obtain the solution of \eqref{sub1}

\begin{equation}\label{up u}
\mathbf{u}^{k+1}=\frac{\rho}{\rho+1}\left(\mathbf{x}^k-\mathbf{x}_o+\frac{\mathbf{y}^k}{\rho}\right).
\end{equation}

The second subproblem \eqref{sub2} can be equivalently written as

\begin{equation}\label{2sub2model}
\begin{aligned}
& \underset{\mathbf{x} \in \mathbb{C}^{\ell N}}{\operatorname{minimize}}\left\|\mathbf{x}-\mathbf{b}^k\right\|_2^2, \\
& \text { subject to }\|\mathbf{x}\|_{\infty} \leq \beta,
\end{aligned}
\end{equation}
where $\mathbf{b}^k=\mathbf{u}^{k+1}+\mathbf{x}_o+\frac{\mathbf{y}^k}{\rho}$.

Since the constraint in subproblem \eqref{2sub2model} is a convex constraint, the optimal solution to problem \eqref{2sub2model} satisfy $\Pi_{\mathrm{x} \in \mathcal{X}}\left(\nabla_{\mathbf{x}} L_\rho\left(\mathbf{u}^{k+1}, \mathbf{x}, \mathbf{y}^k\right)=0\right)$, i.e.,

\begin{equation}
\nabla_{\mathbf{x}} L_\rho\left(\mathbf{u}^{k+1}, \mathbf{x}, \mathbf{y}^k\right)=\mathbf{x}-\mathbf{b}^k.
\end{equation}

Then, we can obtain the solution of \eqref{sub2}

\begin{equation}
\mathbf{x}^{k+1}=\Pi_{\mathbf{x} \in \mathcal{X}}\left(\mathbf{b}^k\right),
\end{equation}
where $\Pi_{\mathbf{x} \in \mathcal{X}}$ denotes projection onto the $\ell_{\infty}$-norm ball which can be expressed as

\begin{equation}\label{up x1}
x(n)= \begin{cases}\beta e^{j\angle x(n)}, & |x(n)|>\beta, \\ x(n), & |x(n)| \leq \beta.\end{cases}
\end{equation}

We summarize the proposed T-ADMM algorithm for model \eqref{model} in Algorithm \ref{T-ADMM}.
\begin{algorithm}
\caption{The T-ADMM algorithm}
\label{T-ADMM}
\begin{tabular}{l}
 {\bf Initialization:} Initialize $\left(\mathbf{u}^1, \mathbf{x}^1, \mathbf{y}^1\right)$. Set parameters $\left(\beta,\rho\right)$.\\Based on OFDM scheme, set $\mathbf{x}_o$.\\
 {\bf Iterate:} for $k = 1,2,...$ \\
  \hspace{0.2cm} S.1 Solve the subproblem \eqref{sub1}.\\
  \hspace{0.4cm} 1.1 Compute $\mathbf{u}^{k+1}$ via \eqref{up u}.\\
  \hspace{0.2cm} S.2 Solve the subproblem \eqref{sub2}.\\
  \hspace{0.4cm} 2.1 Compute $\mathbf{b}^k=\mathbf{u}^{k+1}+\mathbf{x}_o+\frac{\mathbf{y}^k}{\rho}$,\\
  \hspace{0.4cm} 2.2 Compute $x(n)$ via \eqref{up x1}.\\
  \hspace{0.2cm} S.2 Update Lagrangian multipliers.\\
  \hspace{0.4cm} Compute $\mathbf{y}^{k+1}=\mathbf{y}^k+\rho\left(\mathbf{x}^{k+1}-\mathbf{x}_o-\mathbf{u}^{k+1}\right)$.\\
  {\bf Until} reach the stop-criterion. Then output $\mathbf{x}^{k+1}$.\\
\end{tabular}
\end{algorithm}

\subsection{TCU-ADMM Algorithm Framework}

Since the PAPR constraint in equation \eqref{inf} is a convex approximation of the original non-convex PAPR constraint, the ADMM algorithm might not always achieve the desired PAPR value after solving the problem. In this subsection, we propose an improved algorithm in which the constraint \eqref{inf} is updated in each iteration, instead of keeping it fixed throughout the optimization process. At each iteration of the ADMM algorithm, instead of relying on the fixed threshold $\beta$ in the infinity norm constraint, we adjust $\beta^k$ based on the current value of $\mathbf{x}$ and the target PAPR, which can be expressed as

\begin{equation}\label{proj}
\beta^{k+1}=\alpha\sqrt{\frac{1}{\ell N}}\left\|\mathbf{x}^k\right\|_2.
\end{equation}

The projection $\Pi_{\mathbf{x} \in \mathcal{X}}$ can be expressed as

\begin{equation}\label{up x2}
x(n)= \begin{cases}\beta^{k+1} e^{j \angle x(n)}, & |x(n)|>\beta^{k+1}, \\ x(n), & |x(n)| \leq \beta^{k+1}.\end{cases}
\end{equation}

We summarize the proposed TCU-ADMM algorithm in Algorithm \ref{proposed TCU-ADMM algorithm}.

\begin{algorithm}
\caption{The TCU-ADMM algorithm}
\label{proposed TCU-ADMM algorithm}
\begin{tabular}{l}
 {\bf Initialization:} Initialize $\left(\mathbf{u}^1, \mathbf{x}^1, \mathbf{y}^1\right)$.Choose parameters $\left(\beta,\rho\right)$.\\Based on OFDM scheme, set $\mathbf{x}_o$.\\
 {\bf Iterate:} for $k = 1,2,...$ \\
  \hspace{0.2cm} S.1 Solve the subproblem \eqref{sub1} like T-ADMM.\\
  \hspace{0.2cm} S.2 Solve the subproblem \eqref{sub2}.\\
  \hspace{0.4cm} 2.1 Compute $\mathbf{b}^k=\mathbf{u}^{k+1}+\mathbf{x}_o+\frac{\mathbf{y}^k}{\rho}$,\\
  \hspace{0.4cm} 2.2 Updata $\beta$ via \eqref{proj},\\
  \hspace{0.4cm} 2.3 Compute $x(n)$ via \eqref{up x2}.\\
  \hspace{0.2cm} S.2 Update Lagrangian multipliers like T-ADMM.\\
  {\bf Until} reach the stop-criterion. Then output $\mathbf{x}^{k+1}$.\\
\end{tabular}
\end{algorithm}

\subsection{Performance Analysis}
\subsubsection{Computational Complexity}

First, we analyze the complexity of the T-ADMM algorithm. In Step S.1 of \ref{T-ADMM}, only scalar and vector multiplications, as well as vector additions, are involved. Therefore, the computational complexity of Step S.1 is roughly $\mathcal{O}(\ell N)$. In Step S.2, more specifically, in step 2.1, calculating $\mathbf{b}^k$ also only requires scalar and vector multiplications and additions, with a roughly complexity of $\mathcal{O}(\ell N)$. In step 2.2, the projection operation has a roughly complexity of $\mathcal{O}(\ell N)$. Thus, the overall complexity of Step S.2 is roughly $\mathcal{O}(2\ell N)$. In Step S.3, calculating $\mathbf{y}^{k+1}$ again only involves scalar and vector multiplications and vector additions, leading to a roughly computational complexity of $\mathcal{O}(\ell N)$. Summing the complexity of these steps, we conclude that the overall computational complexity of each iteration of the T-ADMM algorithm is roughly $\mathcal{O}(3\ell N)$.

For the TCU-ADMM algorithm, the only difference from T-ADMM lies in solving the second subproblem, where TCU-ADMM requires updating $\beta^{k+1}$. The computational complexity of this update is also roughly $\mathcal{O}(\ell N)$. Therefore, the overall computational complexity per iteration of the TCU-ADMM algorithm is also roughly $\mathcal{O}(4\ell N)$.

In summary, the complexity of T-ADMM is roughly $\mathcal{O}(\ell N)$ and TCU-ADMM algorithms is roughly $\mathcal{O}(\ell N)$.

\subsubsection{Convergence}
First, we present Lemmas 1-2 and their proofs in Appendix A. Based on these lemmas, we show that if proper parameter is chosen, the augmented Lagrangian function $L(\cdot)$ is sufficient descent in each iteration and is lower-bounded, which leads $L(\cdot)$ to converge as $k \rightarrow \infty$. Then, we have the following theorem to show the convergence properties of the proposed T-ADMM algorithm. Its proof is shown in Appendix B.

\begin{theorem}
Let $\left(\mathbf{u}^k, \mathbf{x}^k, \mathbf{y}^k\right)$ be the sequence generated by the proposed T-ADMM algorithm. If penalty parameter $\rho$ satisfies $\rho\geq 2$, we have the following convergence results

\begin{equation}\label{theorem}
    \begin{aligned}
& \lim _{k \rightarrow \infty}\mathbf{u}^{k+1}=\mathbf{u}^*,\lim _{k \rightarrow \infty}\mathbf{x}^{k+1}=\mathbf{x}^*, \\
&  \lim _{k \rightarrow \infty}\mathbf{y}^{k+1}=\mathbf{y}^*, \lim _{k \rightarrow \infty}\mathbf{x}^{k+1}-\mathbf{x}_o-\mathbf{u}^{k+1}=0.\\
\end{aligned}
\end{equation}
\end{theorem}

{\it Remark:} Theorem 2 proves that the proposed T-ADMM algorithm converges if proper parameter is chosen. However, for the TCU-ADMM algorithm, the constraint update \eqref{proj} is heuristic, making the theoretical proof of its convergence even more challenging. Nevertheless, simulations presented in this paper demonstrate that both the T-ADMM and TCU-ADMM algorithms converge consistently.
\section{SIMULATION RESULTS}
In this section, we present simulation results to evaluate the performance of the proposed T-ADMM and TCU-ADMM methods. We compare the proposed algorithms with the ICF method\cite{12} and the ADMM-Direct method\cite{22}.

The simulation parameters are as follows: we consider an OFDM signal with subcarriers $N=512$ and set the oversampling rate $\ell=4$\cite{30}. The PAPR is set to 4 dB. The modulation scheme is set to QPSK and 16QAM respectively. In the simulations, we randomly generate 5000 symbols.

\begin{figure}[h]
\centering
  \centerline{\includegraphics[scale=0.45]{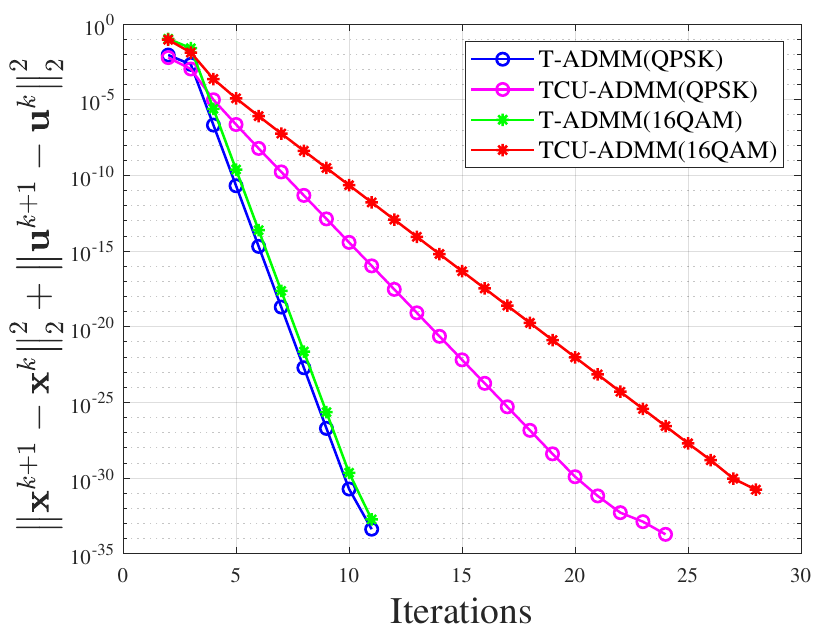}}
\caption{ The convergence performance of T-ADMM and TCU-ADMM with QPSK and 16QAM modulation.}
\label{converg TADMM}
\end{figure}

First, we conduct a convergence analysis for proposed T-ADMM and TCU-ADMM. Figure \ref{converg TADMM} shows the convergence curves of the PAPR reduction model, illustrating the impact of the iteration number $k$ on the residuals $\left\|\mathbf{x}^{k+1}-\mathbf{x}^k\right\|_2^2+\left\|\mathbf{u}^{k+1}-\mathbf{u}^k\right\|_2^2$. From the curves, we can see that as the number of iterations increases, the residuals show a downward trend, indicating that both ADMM algorithms converge. Furthermore, the TCU-ADMM algorithm demonstrates a faster reduction in residuals compared to the T-ADMM algorithm. While we do not provide an exact proof of the convergence of the TCU-ADMM algorithm, our observations from Figure \ref{converg TADMM} suggest that the TCU-ADMM indeed converges.

\begin{table}[t]
\renewcommand \arraystretch{1.5}
\caption{COMPLEXITY COMPARISON OF DIFFERENT ALGORITHM}
\label{Complexity}
\centering
\begin{tabular}{c|c}
\hline
Algorithm   & Computational Complexity \\\hline
T-ADMM      &$\mathcal{O}\left(3\ell N\right)$\\\hline
TCU-ADMM    &$\mathcal{O}\left(4\ell N\right)$\\\hline
ADMM-Direct &$\mathcal{O}\left(3\ell N \log _2 \ell N\right)$\\\hline
ICF         &$\mathcal{O}\left(2\ell N \log _2 \ell N\right)$\\\hline                         \end{tabular}
\end{table}

Table \ref{Complexity} compares the computational complexity of different algorithms. Specifically, In contrast, the complexities of T-ADMM and TCU-ADMM are $\mathcal{O}\left(3\ell N\right)$ and $\mathcal{O}\left(4\ell N\right)$, significantly lower than other methods. These results demonstrate that the proposed algorithms effectively reduce computational complexity, particularly T-ADMM and TCU-ADMM, which exhibit superior low-complexity characteristics, making them well-suited for practical system applications.

\begin{figure*}[tp!]
\centering
\begin{minipage}{0.3\linewidth}
  \includegraphics[scale=0.4]{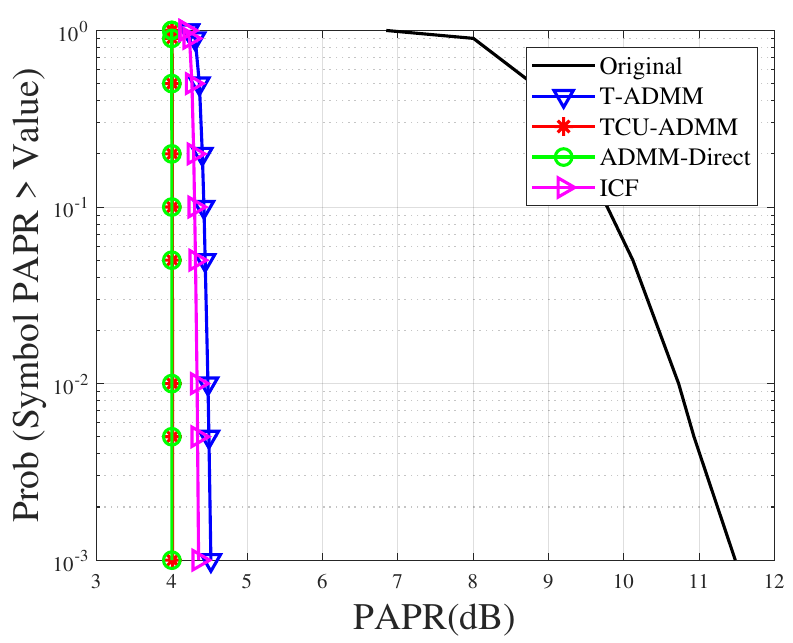}
\centerline{(a) PAPR reduction performance }
\end{minipage}
\begin{minipage}{0.3\linewidth}
  \includegraphics[scale=0.4]{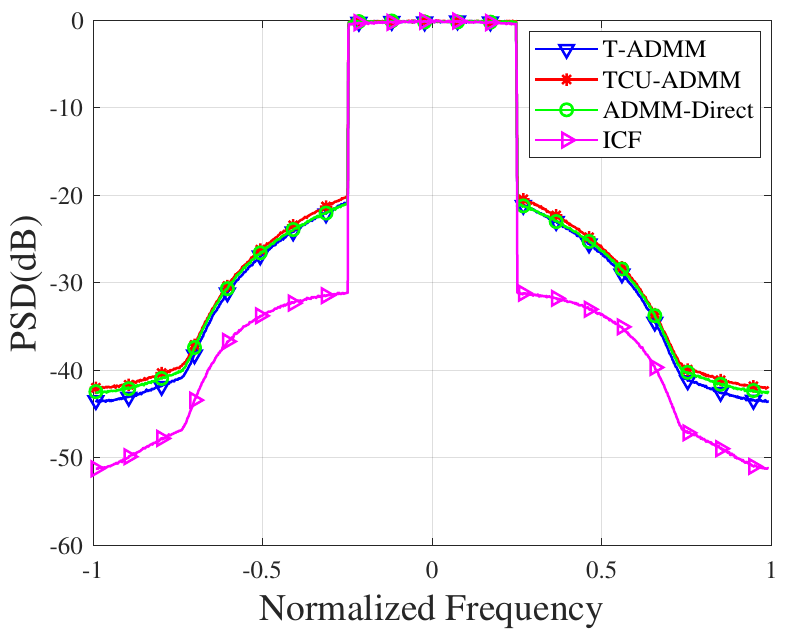}
  \centerline{(b) OOBE performance}
  \end{minipage}
  \begin{minipage}{0.3\linewidth}
  \includegraphics[scale=0.4]{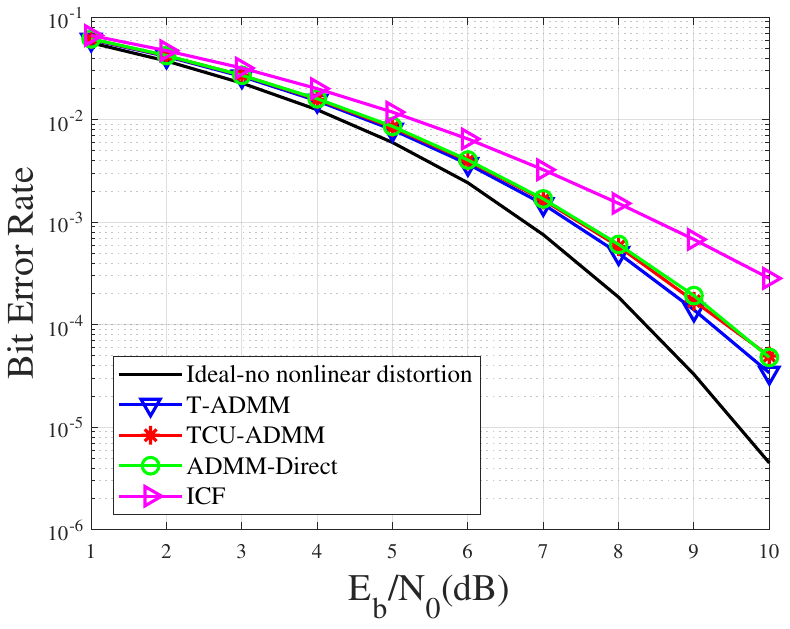}
  \centerline{(c) BER performance}
  \end{minipage}
\caption{ The performance of different methods with QPSK modulation.}
\label{qpsk}
\end{figure*}

\begin{figure*}[tp!]
\centering
\begin{minipage}{0.3\linewidth}
  \includegraphics[scale=0.4]{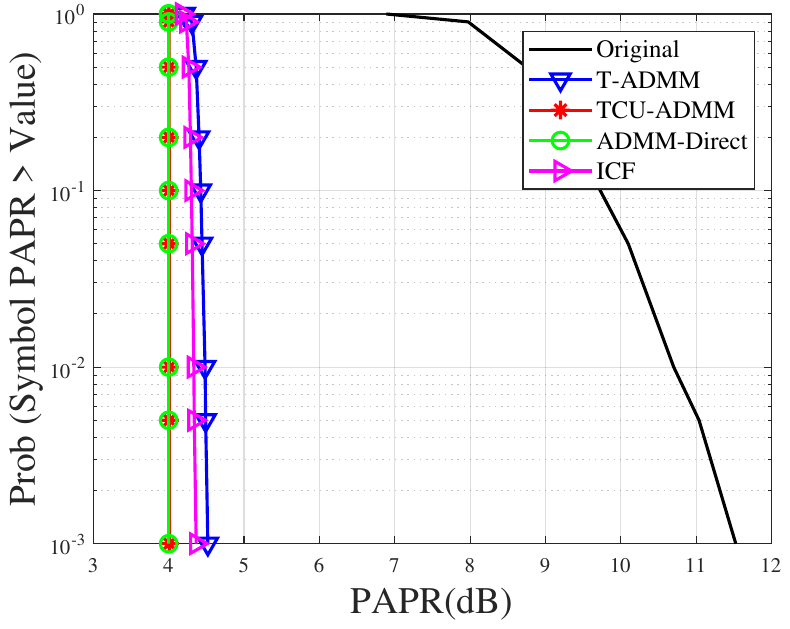}
\centerline{(a) PAPR reduction performance }
\end{minipage}
\begin{minipage}{0.3\linewidth}
  \includegraphics[scale=0.4]{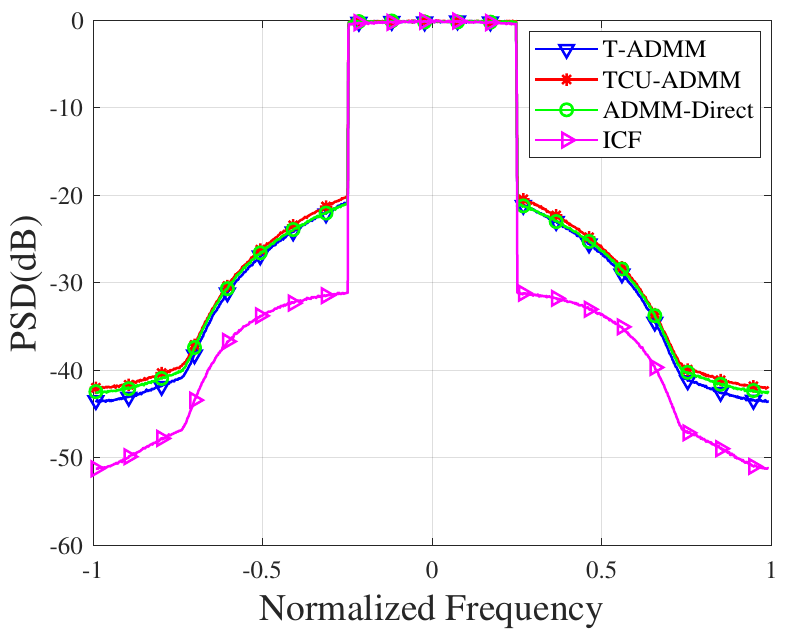}
  \centerline{(b) OOBE performance}
  \end{minipage}
  \begin{minipage}{0.3\linewidth}
  \includegraphics[scale=0.4]{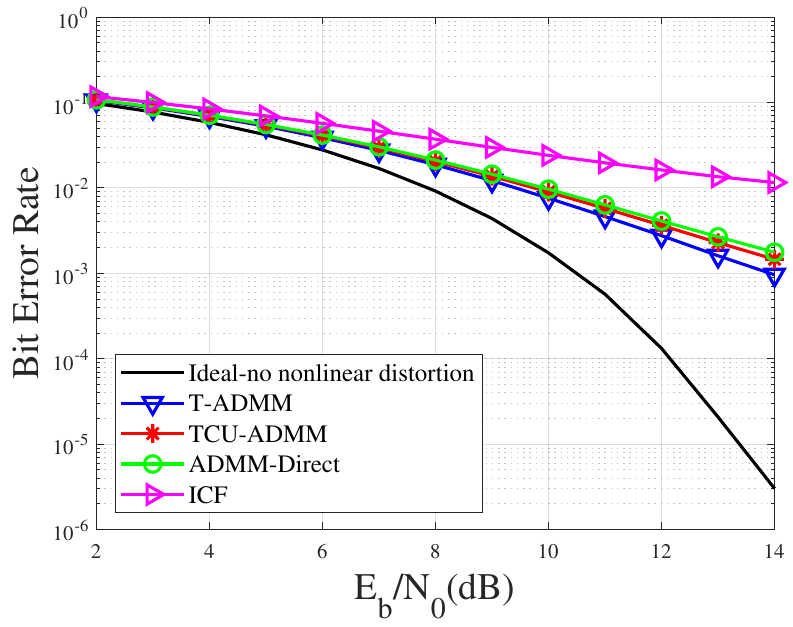}
  \centerline{(c) BER performance}
  \end{minipage}
\caption{ The performance of different methods with 16QAM modulation.}
\label{16qam}
\end{figure*}

Figures \ref{qpsk} and \ref{16qam} present a comprehensive evaluation of the PAPR complementary cumulative distribution function (CCDF), OOBE, and bit error rate (BER) performance for both the original OFDM signal and the PAPR-reduced OFDM signal after passing through a soft-limiter solid-state power amplifier (SSPA) with a smoothing factor of 3 \cite{24} in an additive white Gaussian noise (AWGN) channel. The simulations for T-ADMM and TCU-ADMM algorithms was conducted over 5 iterations. The CCDF represents the probability that the PAPR of the OFDM symbols exceeds a given PAPR threshold $\mathrm{PAPR}_0$, expressed as $\mathrm{CCDF}=\operatorname{Pr}\left(\mathrm{PAPR}>\mathrm{PAPR}_0\right)$.

The CCDF metric, depicted in Figures \ref{qpsk}(a) and \ref{16qam}(a), represents the probability that the PAPR of the OFDM symbols exceeds a predefined threshold $\mathrm{PAPR}_0$. In Figure \ref{qpsk}(a), the ``Original'' curve represents the CCDF of undistorted OFDM signals. The results show that the proposed TCU-ADMM algorithm achieves CCDF performance comparable to ADMM-Direct, exhibiting a sharp cutoff in their CCDF curves. In contrast, the CCDF curves for the ICF algorithm and the proposed T-ADMM method are shifted to the right, indicating higher residual PAPR values. This demonstrates that the proposed algorithms achieves effective PAPR reduction.

The performance of OOBE is depicted in Figures \ref{qpsk}(b) and \ref{16qam}(b). The results indicate that T-ADMM and TCU-ADMM maintain OOBE levels similar to ADMM-Direct. The ICF algorithm, which employs filtering before the SSPA, shows no out-of-band radiation prior to entering the SSPA, resulting in its low OOBE levels.

Figure \ref{qpsk}(c) and \ref{16qam}(c) illustrates the BER performance of the optimized OFDM symbols after passing through the SSPA, with an input power back-off (IBO) from saturation set to 4.1 dB. The ``Ideal'' curve represents an OFDM system without distortion. T-ADMM and TCU-ADMM achieve BER performance comparable to ADMM-Direct.

Overall, the proposed T-ADMM and TCU-ADMM algorithm demonstrate comparable performance to ADMM-Direct, with much lower computational complexity.

\section{CONCLUSION}
In this paper, we propose two FFT-free PAPR reduction algorithms for OFDM signals, T-ADMM and TCU-ADMM. These algorithms have linear computational complexity. The proposed algorithms can yield the desired OFDM symbols after just a few iterations  while achieving significant PAPR reduction and maintaining good OOBE and BER performance. Additionally, we prove that the T-ADMM algorithm is theoretically guaranteed convergent if proper parameter is chosen. Compared to existing algorithms, the proposed methods address the issue of excessive complexity in PAPR reduction for OFDM.

\appendices

\section{PROOF LEMMA 1}
\begin{lemma}
The following inequality exist

\begin{equation}
 \begin{aligned}
& L\left(\mathbf{u}^k, \mathbf{x}^k, \mathbf{y}^k\right)-L\left(\mathbf{u}^{k+1}, \mathbf{x}^{k+1}, \mathbf{y}^{k+1}\right) \\
& \geq \frac{\rho}{2}\left\|\mathbf{u}^k-\mathbf{u}^{k+1}\right\|_2^2+\frac{\rho}{2}\left\|\mathbf{x}^k-\mathbf{x}^{k+1}\right\|_2^2+\frac{1}{\rho}\left\|\mathbf{y}^k-\mathbf{y}^{k+1}\right\|_2^2.
\end{aligned}
\end{equation}

\end{lemma}

{\it Proof:} To facilitate the subsequent derivations,we define the following quantities

\begin{equation}
\begin{aligned}
&\Delta_{\mathbf{u}}^k=L\left(\mathbf{u}^k, \mathbf{x}^k, \mathbf{y}^k\right)-L\left(\mathbf{u}^{k+1}, \mathbf{x}^k, \mathbf{y}^k\right), \\
&\Delta_{\mathbf{x}}^k=L\left(\mathbf{u}^{k+1}, \mathbf{x}^k, \mathbf{y}^k\right)-L\left(\mathbf{u}^{k+1}, \mathbf{x}^{k+1}, \mathbf{y}^k\right), \\
&\Delta_y^k=L\left(\mathbf{u}^{k+1}, \mathbf{x}^{k+1}, \mathbf{y}^k\right)-L\left(\mathbf{u}^{k+1}, \mathbf{x}^{k+1}, \mathbf{y}^{k+1}\right).
\end{aligned}
\end{equation}

By summing these terms, we have

\begin{equation}\label{cancha}
    \begin{aligned}
& L\left(\mathbf{u}^k, \mathbf{x}^k, \mathbf{y}^k\right)-L\left(\mathbf{u}^{k+1}, \mathbf{x}^{k+1}, \mathbf{y}^{k+1}\right) \\
& =\Delta_{\mathbf{u}}^k+\Delta_{\mathbf{x}}^k+\Delta_{\mathbf{y}}^k.
\end{aligned}
\end{equation}

Since $L\left(\mathbf{u}, \mathbf{x}^k, \mathbf{y}^k\right)$ is strongly convex with respect to $\mathbf{u}$, we can bound $\Delta_{\mathbf{u}}^k$ using the quadratic property of strong convexity\cite{26}

\begin{equation}
    \Delta_{\mathbf{u}}^k \geq\left\langle\nabla_{\mathbf{u}} L\left(\mathbf{u}^{k+1}, \mathbf{x}^k, \mathbf{y}^k\right), \mathbf{u}^k-\mathbf{u}^{k+1}\right\rangle+\frac{\rho}{2}\left\|\mathbf{u}^k-\mathbf{u}^{k+1}\right\|_2^2.
\end{equation}
Since $\mathbf{u}^{k+1}=\arg \min L\left(\mathbf{u}, \mathbf{x}^k, \mathbf{y}^k\right)$, we have

\begin{equation}
    \nabla_{\mathbf{u}} L\left(\mathbf{u}^{k+1}, \mathbf{x}^k, \mathbf{y}^k\right)=0.
\end{equation}

Thus,

\begin{equation}
    \Delta_{\mathbf{u}}^k \geq \frac{\rho}{2}\left\|\mathbf{u}^k-\mathbf{u}^{k+1}\right\|_{2}^{2}.
\end{equation}

For $\Delta_{\mathbf{x}}^k$, since $L\left(\mathbf{u}^{k+1}, \mathbf{x}, \mathbf{y}^k\right)$ is strongly convex with respect to $\mathbf{x}$, we can also bound $\Delta_{\mathbf{x}}^k$ using the quadratic property of strong convexity

\begin{equation}
    \Delta_{\mathbf{x}}^k \geq\left\langle\nabla_{\mathbf{x}} L\left(\mathbf{u}^{k+1}, \mathbf{x}^{k+1}, \mathbf{y}^k\right), \mathbf{x}^k-\mathbf{x}^{k+1}\right\rangle+\frac{\rho}{2}\left\|\mathbf{x}^k-\mathbf{x}^{k+1}\right\|_2^2.
\end{equation}

 Since subproblem \eqref{2sub2model} is a convex problem with a convex constraint, $\mathbf{x}^{k+1}=\Pi_{\mathrm{x} \in \mathcal{X}}\left(\nabla_{\mathbf{x}} L_\rho\left(\mathbf{u}^{k+1}, \mathbf{x}, \mathbf{y}^k\right)=0\right)$ is the optimal solution of subproblem \eqref{2sub2model}, we have

\begin{equation}
    \left\langle\nabla_{\mathbf{x}} L\left(\mathbf{u}^{k+1}, \mathbf{x}^{k+1}, \mathbf{y}^k\right), \mathbf{x}-\mathbf{x}^{k+1}\right\rangle\geq 0.
\end{equation}

Then,

\begin{equation}
    \left\langle\nabla_{\mathbf{x}} L\left(\mathbf{u}^{k+1}, \mathbf{x}^{k+1}, \mathbf{y}^k\right), \mathbf{x}^k-\mathbf{x}^{k+1}\right\rangle\geq 0.
\end{equation}

Thus,

\begin{equation}
    \Delta_{\mathbf{x}}^k \geq \frac{\rho}{2}\left\|\mathbf{x}^k-\mathbf{x}^{k+1}\right\|_{2}^{2}.
\end{equation}

For $\Delta_{\mathbf{y}}^k$, plugging \eqref{lag} into \eqref{cancha}, we have the following equality

\begin{equation}
\Delta_{\mathbf{y}}^k=\frac{1}{\rho}\left\|\mathbf{y}^k-\mathbf{y}^{k+1}\right\|_2^2 .
\end{equation}

Combining $\Delta_{\mathbf{u}}^k$, $\Delta_{\mathbf{x}}^k$ and $\Delta_{\mathbf{y}}^k$, we obtain

\begin{equation}
 \begin{aligned}
& L\left(\mathbf{u}^k, \mathbf{x}^k, \mathbf{y}^k\right)-L\left(\mathbf{u}^{k+1}, \mathbf{x}^{k+1}, \mathbf{y}^{k+1}\right) \\
& \geq \frac{\rho}{2}\left\|\mathbf{u}^k-\mathbf{u}^{k+1}\right\|_2^2+\frac{\rho}{2}\left\|\mathbf{x}^k-\mathbf{x}^{k+1}\right\|_2^2+\frac{1}{\rho}\left\|\mathbf{y}^k-\mathbf{y}^{k+1}\right\|_2^2.
\end{aligned}
\end{equation}
$\hfill\blacksquare$

\section{PROOF LEMMA 2}
\begin{lemma}
If $\rho \geq 2$, the following inequality exist

\begin{equation}
L\left(\mathbf{u}^{k+1}, \mathbf{x}^{k+1}, \mathbf{y}^{k+1}\right) \geq 0 .
\end{equation}

\end{lemma}

{\it Proof:} Let $f\left(\mathbf{u}^{k+1}\right)=\frac{1}{2}\left\|\mathbf{u}^{k+1}\right\|_2^2$. Since $\mathbf{y}^{k+1}=\nabla f\left(\mathbf{u}^{k+1}\right)$, we can rewrite $L_\rho$ as

\begin{equation}
\begin{aligned}
&L_\rho\left(\mathbf{u}^{k+1}, \mathbf{x}^{k+1}, \mathbf{y}^{k+1}\right)\\
&=f\left(\mathbf{u}^{k+1}\right)+\left\langle\nabla f\left(\mathbf{u}^{k+1}\right), \mathbf{x}^{k+1}-\mathbf{x}_o-\mathbf{u}^{k+1}\right\rangle\\
&+\frac{\rho}{2}\left\|\mathbf{x}^{k+1}-\mathbf{x}_o-\mathbf{u}^{k+1}\right\|_2^2 .
\end{aligned}
\end{equation}

Since $f\left(\mathbf{u}^{k+1}\right)$ is Lipschitz continuous, we have

\begin{equation}
\left\|\nabla f\left(\mathbf{x}^{k+1}-\mathbf{x}_o\right)-\nabla f\left(\mathbf{u}^{k+1}\right)\right\| \leq 2\left\|\mathbf{x}^{k+1}-\mathbf{x}_o-\mathbf{u}^{k+1}\right\|_2.
\end{equation}

Thus, the term $\left\langle\nabla f\left(\mathbf{u}^{k+1}\right), \mathbf{x}^{k+1}-\mathbf{x}_o-\mathbf{u}^{k+1}\right\rangle$ can be bounded as

\begin{equation}
\begin{aligned}
&\left\langle\nabla f\left(\mathbf{u}^{k+1}\right), \mathbf{x}^{k+1}-\mathbf{x}_o-\mathbf{u}^{k+1}\right\rangle\\
&\geq  \left\langle\nabla f\left(\mathbf{x}^{k+1}-\mathbf{x}_o\right), \mathbf{x}^{k+1}-\mathbf{x}_o-\mathbf{u}^{k+1}\right\rangle \\
& -\left\|\mathbf{x}^{k+1}-\mathbf{x}_o-\mathbf{u}^{k+1}\right\|_2^2.
\end{aligned}
\end{equation}

Substituting this into $L_\rho$, we get

\begin{equation}
 \begin{aligned}
&L_\rho\left(\mathbf{u}^{k+1}, \mathbf{x}^{k+1}, \mathbf{y}^{k+1}\right) \\
&\geq  f\left(\mathbf{u}^{k+1}\right)+\left\langle\nabla f\left(\mathbf{x}^{k+1}-\mathbf{x}_o\right), \mathbf{x}^{k+1}-\mathbf{x}_o-\mathbf{u}^{k+1}\right\rangle \\
& +\left(\frac{\rho}{2}-1\right)\left\|\mathbf{x}^{k+1}-\mathbf{x}_o-\mathbf{u}^{k+1}\right\|_2^2.
\end{aligned}
\end{equation}

Since $\nabla f\left(\mathbf{x}^{k+1}-\mathbf{x}_o\right)$ is also Lipschitz continuous, we further have

\begin{equation}
 \begin{aligned}
&f\left(\mathbf{x}^{k+1}-\mathbf{x}_o\right) \\
&\leq  f\left(\mathbf{u}^{k+1}\right)+\left\langle\nabla f\left(\mathbf{x}^{k+1}-\mathbf{x}_o\right), \mathbf{x}^{k+1}-\mathbf{x}_o-\mathbf{u}^{k+1}\right\rangle \\
& +\frac{1}{2}\left\|\mathbf{x}^{k+1}-\mathbf{x}_o-\mathbf{u}^{k+1}\right\|_2^2.
\end{aligned}
\end{equation}

Therefore,

\begin{equation}\label{lemma2}
\begin{aligned}
&L_\rho\left(\mathbf{u}^{k+1}, \mathbf{x}^{k+1}, \mathbf{y}^{k+1}\right) \\
&\geq f\left(\mathbf{x}^{k+1}-\mathbf{x}_o\right)+\left(\frac{\rho}{2}-1\right)\left\|\mathbf{x}^{k+1}-\mathbf{x}_o-\mathbf{u}^{k+1}\right\|_2^2 .
\end{aligned}
\end{equation}

If $\rho \geq 2$, $\forall k, L_\rho\left(\mathbf{u}^{k+1}, \mathbf{x}^{k+1}, \mathbf{y}^{k+1}\right)\geq 0 $. $\hfill\blacksquare$

\section{PROOF THEOREM 2}

{\it Proof:} First of all, we set the parameter $\rho$ satisfies the condition in Lemma 2. Summing both sides of the inequality \eqref{lemma2} at $k = 1,2,...,+\infty$, we can obtain

\begin{equation}
    \begin{aligned}
& L_\rho\left(\mathbf{u}^1, \mathbf{x}^1, \mathbf{y}^1\right)-L_\rho\left(\mathbf{u}^{k+1}, \mathbf{x}^{k+1}, \mathbf{y}^{k+1}\right) \\
& \geq \sum_{k=1}^{\infty} \frac{\rho}{2}\left\|\mathbf{u}^k-\mathbf{u}^{k+1}\right\|_2^2+\frac{\rho}{2}\left\|\mathbf{x}^k-\mathbf{x}^{k+1}\right\|_2^2+\frac{1}{\rho}\left\|\mathbf{y}^k-\mathbf{y}^{k+1}\right\|_2^2.
\end{aligned}
\end{equation}

Since $\rho>0$ and $L_\rho\left(\mathbf{u}^1, \mathbf{x}^1, \mathbf{y}^1\right)$ is bounded, we conclude that

\begin{subequations}\label{converge}
    \begin{align}
& \lim _{k \rightarrow \infty}\left\|\mathbf{u}^k-\mathbf{u}^{k+1}\right\|_2^2=0, \\
& \lim _{k \rightarrow \infty}\left\|\mathbf{x}^k-\mathbf{x}^{k+1}\right\|_2^2=0, \\
& \lim _{k \rightarrow \infty}\left\|\mathbf{y}^k-\mathbf{y}^{k+1}\right\|_2^2=0.\label{converge1}
\end{align}
\end{subequations}

Plugging \eqref{converge1} into \eqref{lag} we further have

\begin{equation}\label{converge2}
\lim _{k \rightarrow \infty}\left\|\mathbf{x}^{k+1}-\mathbf{x}_o-\mathbf{u}^{k+1}\right\|_2^2=0. \end{equation}

Moreover, combining \eqref{converge} and \eqref{converge2}, and then letting $\left(\mathbf{u}^*, \mathbf{x}^*, \mathbf{y}^*\right)$ be the solution of the proposed T-ADMM algorithm, we have

\begin{equation}
    \begin{aligned}
& \lim _{k \rightarrow \infty}\mathbf{u}^{k+1}=\mathbf{u}^*,\lim _{k \rightarrow \infty}\mathbf{x}^{k+1}=\mathbf{x}^*, \\
&  \lim _{k \rightarrow \infty}\mathbf{y}^{k+1}=\mathbf{y}^*, \lim _{k \rightarrow \infty}\mathbf{x}^{k+1}-\mathbf{x}_o-\mathbf{u}^{k+1}=0,\\
\end{aligned}
\end{equation}
$\hfill\blacksquare$


\ifCLASSOPTIONcaptionsoff
  \newpage
\fi


\begin{thebibliography}{00}
\bibitem{1}A. Ghosh, R. Ratasuk, B. Mondal, N. Mangalvedhe and T. Thomas, ``LTE-advanced: next-generation wireless broadband technology [Invited Paper]," \emph{IEEE Wireless Commun.}, vol. 17, no. 3, pp. 10-22, June 2010.
\bibitem{2}  A. M. Jaradat, J. M. Hamamreh, and H. Arslan, ``Modulation options for OFDM-based waveforms: Classification, comparison, and future directions," \emph{IEEE Access}, vol. 7, pp. 17263-17278, 2019.
\bibitem{3}  S. H. Han and J. H. Lee, ``An overview of peak-to-average power ratio reduction techniques for multicarrier transmission," \emph{IEEE Wireless Commun.}, vol. 12, no. 2, pp. 56-65, Apr. 2005.
\bibitem{24}	Y. Rahmatallah and S. Mohan, ``Peak-To-Average Power Ratio Reduction in OFDM Systems: A Survey And Taxonomy," \emph{IEEE Commun. Surveys Tut.}, vol. 15, no. 4, pp. 1567-1592, Fourth Quarter 2013.
\bibitem{5}	J. A. Davis and J. Jedwab, ``Peak-to-mean power control in OFDM, Golay complementary sequences, and Reed-Muller codes," \emph{IEEE Trans. Inf. Theory}, vol. 45, no. 7, pp. 2397-2417, 1999.
 \bibitem{6}S. Shu, D. Qu, L. Li, and T. Jiang, ``Invertible subset QC-LDPC codes for PAPR reduction of OFDM signals,"  \emph{IEEE Trans. Broadcast.}, vol. 61, no. 2, pp. 290-298, Jun. 2015.
\bibitem{7}	M. Lin, K. Chen and S. Li, ``Turbo coded OFDM system with peak power reduction," \emph{2003 IEEE 58th Vehicular Technology Conference}, Vol. 4, pp. 2282-2286, 2003.
\bibitem{8}	N. Taspinar and M. Yildirim, ``A novel parallel artificial bee colony algorithm and its PAPR reduction performance using SLM scheme in OFDM and MIMO-OFDM systems," \emph{IEEE Commun. Lett.}, vol 19, no. 10, pp. 1830-1833, 2015.
\bibitem{9}	Y. J. Cho, K. H. Kim, and J. Y. Woo, ``Low-complexity PTS schemes using dominant time-domain samples in OFDM systems,`` \emph{IEEE Trans. Broadcast.}, vol. 63, no. 2, pp. 440-445, 2017.
\bibitem{12}	J. Armstrong, ``Peak-to-average power reduction for OFDM by repeated clipping and frequency domain filtering," \emph{Electron. Lett.}, vol. 38, no. 5, pp. 246-247, 2002.
\bibitem{13}  M.-X. Hu, Y.-Z. Li, W. Wang, and H.-L. Zhang, ``A piecewise linear companding transform for PAPR reduction of OFDM signals with companding distortion mitigation," \emph{IEEE Trans. Broadcast.}, vol. 60, no.3, pp. 532-539, Sep. 2014.
\bibitem{15}	A. Aggarwal and T. Meng, ``Minimizing the peak-to-average power ratio of OFDM signals using convex optimization," \emph{IEEE Trans. Signal Process.}, vol. 54, no. 8, pp. 3099-3110, Aug. 2006.
\bibitem{16}	Q. Liu, R. J. Baxley, X. Ma, and G. T. Zhou, ``Error vector magnitude optimization for OFDM systems with a deterministic peak-to-average power ratio constraint," \emph{IEEE J. Select. Topics Signal Process.}, vol. 3, no. 3, pp. 418-429, Jun. 2009.
\bibitem{17}	Y.-C. Wang and Z.-Q. Luo, ``Optimized iterative clipping and filtering for PAPR reduction of OFDM signals," \emph{IEEE  Trans. Commun.}, vol. 59, no. 1, pp. 33-37, Jan. 2011.
\bibitem{18}	Y. -C. Wang, J. -L. Wang, K. -C. Yi and B. Tian, ``PAPR Reduction of OFDM Signals With Minimized EVM via Semidefinite Relaxation," \emph{IEEE Trans. Veh. Technol.}, vol. 60, no. 9, pp. 4662-4667, Nov. 2011.
\bibitem{20}	S. Boyd, N. Parikh, E. Chu, B. Peleato, and J. Eckstein, ``Distributed optimization and statistical learning via the alternating direction method of multipliers," \emph{Found. Trends Mach. Learn.}, vol. 3, no. 1, pp. 1-122, Jan. 2011.
\bibitem{21}	H. Bao, J. Fang, Q. Wan, Z. Chen and T. Jiang, ``An ADMM Approach for PAPR Reduction for Large-Scale MIMO-OFDM Systems," \emph{IEEE Trans. Veh. Technol.}, vol. 67, no. 8, pp. 7407-7418, Aug. 2018.
\bibitem{22}	Y. Wang, Y. Wang and Q. Shi, ``Optimized Signal Distortion for PAPR Reduction of OFDM Signals With IFFT/FFT Complexity Via ADMM Approaches," \emph{IEEE Trans. Signal Process.}, vol. 67, no. 2, pp. 399-414, Jan. 2019.
\bibitem{26}D. P. Bertsekas, \emph{Nonlinear Programming}. Belmont, MA, USA: Athena Scientific, pp. 667-668, 2nd ed., 1999.
\bibitem{30}H. Ochiai and H. Imai, ``On the distribution of the peak-to-average power ratio in OFDM signals," \emph{IEEE Trans. Commun.}, vol. 49, no. 2, pp. 282-289, Feb. 2001.

\end{thebibliography}
\end{document}